# Comet C/2004 Q2 (Machholz): Parent Volatiles, A Search for Deuterated Methane, and Constraint on the $CH_4$ Spin Temperature

Boncho P. Bonev[1,2], Michael J. Mumma[2], Erika L. Gibb[3], Michael A. DiSanti[2], Geronimo L. Villanueva[1,2], Karen Magee-Sauer[4], and Richard S. Ellis[5]

[1] Department of Physics, The Catholic Univ. of America, Washington DC, 20064; bonev@cua.edu
[2] Solar System Exploration Division, Mailstop 693, NASA's Goddard Space Flight Center, Greenbelt, MD 20771
[3] Department of Physics and Astronomy, University of Missouri – St. Louis, St. Louis, MO 63121
[4] Department of Physics and Astronomy, Rowan Univ., Glassboro, NJ, 08028-1701
[5] Department of Astronomy, California Institute of Technology, Mail-Code 105-24, 1200 East California Blvd., Pasadena, CA, 91125

ABSTRACT

High-dispersion ($\lambda/\delta\lambda \approx 25{,}000$) infrared spectra of Comet C/2004 Q2 (Machholz) were acquired on Nov. 28-29, 2004, and Jan. 19, 2005 (UT dates) with NIRSPEC at the Keck-2 telescope on Mauna Kea. We detected $H_2O$, $CH_4$, $C_2H_2$, $C_2H_6$, CO, $H_2CO$, $CH_3OH$, HCN, and $NH_3$ and we conducted a sensitive search for $CH_3D$. We report rotational temperatures, production rates, and mixing ratios (with respect to $H_2O$) at heliocentric distances of 1.49 AU (Nov. 2004) and 1.21 AU (Jan. 2005). We highlight three principal results: (1) The mixing ratios of parent volatiles measured at 1.49 AU and 1.21 AU agree within confidence limits, consistent with homogeneous composition in the mean volatile release from the nucleus of C/2004 Q2. Notably, the relative abundance of $C_2H_6/C_2H_2$ is substantially higher than those measured in other comets, while the mixing ratios $C_2H_6/H_2O$, $CH_3OH/H_2O$, and $HCN/H_2O$ are similar to those observed in comets, referred to as "organics-normal". (2) The spin temperature of $CH_4$ is > 35-38 K, an estimate consistent with the more robust spin temperature found for $H_2O$. (3) We obtained a 3σ upper limit of $CH_3D/CH_4 < 0.020$ (D/H < 0.005). This limit suggests that methane released from the nucleus of C/2004 Q2 is not dominated by a component formed in extremely cold (near 10 K) environments. Formation pathways of both interstellar and nebular origin consistent with the measured D/H in methane are discussed. Evaluating the relative contributions of these pathways requires further modeling of chemistry including both gas-phase and gas-grain processes in the natal interstellar cloud and in the protoplanetary disk.

*Subject headings*: comets: general - comets: individual (C/2004 Q2 [Machholz]) – infrared: solar system – solar system: formation – ISM: molecules

1. ESTABLISHING DIVERSITY IN COMETARY COMPOSITION

The composition of comets provides a major observational constraint in cosmogony. The volatiles and dust stored in comet nuclei preserve a physico-chemical record for the conditions (temperature, pressure, radiation field density) under which cometary material formed (Wooden 2008; DiSanti & Mumma 2008; Bockelée-Morvan et al. 2004; Irvine et al. 2000; Mumma et al. 1993).

Comparing relative abundances of parent[1] molecules ($H_2O$, HCN, CO, etc.) in comets to those found in the interstellar medium (ISM) and in disks around young stars is vital to understanding the different stages of planetary system formation and the processing history experienced by organic matter during the evolutionary transitions between these stages (see Charnley & Rodgers 2008a, 2008b; Carr & Najita 2008; Ehrenfreund et al. 2006; Gibb et al. 2007a; Boogert et al. 2004; Nuth et al. 2000). Measuring the native[1] volatile composition of comets tests dynamical models for the early solar system (Böhnhardt et al. 2008, Kobayashi et al. 2007, Mumma et al. 2005). Studies of comets are also important for assessing the possibility of exogenous delivery of water and pre-biotic organics to early Earth as a hypothesized precursor event(s) leading to development of the biosphere (Delsemme 1998, 2000).

Spectroscopy has revealed a number of molecules released from comet nuclei and a significant diversity in the relative abundances (or "mixing ratios") among them (see DiSanti & Mumma 2008; Crovisier 2006; Dello Russo et al. 2008; Magee-Sauer et al. 2008; Biver et al. 2002; Mumma et al. 2001). This chemical diversity has yet to be fully explored, because the sample of observed comets is fairly small, especially at infrared (IR) wavelengths where molecules with no permanent dipole moment ($CH_4$, $C_2H_2$, $C_2H_6$) can be uniquely sensed.

Moreover, in addition to mixing ratios (e.g., $CH_4/H_2O$, $HCN/H_2O$, $CO/H_2CO/CH_3OH$, etc.), other (possible) cosmogonic signatures such as isotopic (D/H, $^{15}N/^{14}N$) and nuclear-isomeric (ortho-para, A:E:F, etc.) ratios in particular species are of keen interest in cometary science. With few exceptions the knowledge of these important "observables" is very limited (see the reviews of Charnley & Rodgers 2008a; Crovisier 2006). In the case of methane (of particular interest in this work), an upper limit for $CH_3D/CH_4$ equal to 0.04 (95% confidence limit) was found in comet C/2004 Q4 (NEAT) (Kawakita et al. 2005), based on a single spectral line. The same work also reported a spin temperature for $CH_4$ ($T_{spin}$ ~ 33 K, corresponding to the A:E:F nuclear isomeric ratio). Measurements of both deuterium enrichment and spin temperatures of various species in many more comets are critical for testing possible variations among the comet population. The inter-relation of such measurements with other plausible cosmogonic indicators (mixing ratios of parent volatiles, properties of cometary dust) will (likely) clarify the origin of cometary material.

---

[1] "Parent" molecules are those released directly from ices stored in the nucleus ("native" ices).

This paper summarizes the results from our comprehensive study of parent volatiles in comet C/2004 Q2 (Machholz)[2], observed with the Near InfraRed echelle SPECtrograph (NIRSPEC) (McLean et al. 1998) at the W. M. Keck Observatory atop Mauna Kea, Hawaii. We detected a number of parent volatiles and we conducted a sensitive search for $CH_3D$. We highlight three principal results:

1. An upper limit for the deuterium enrichment of methane ($CH_3D/CH_4$).
2. An lower bound for the spin temperature of methane.
3. Mixing ratios of $CH_4$, $C_2H_2$, $C_2H_6$, $H_2CO$, $CH_3OH$, HCN, and $NH_3$ relative to $H_2O$, measured (or stringently constrained) at heliocentric distances of 1.49 AU and 1.21 AU. Mixing ratio $CO/H_2O$ measured at 1.48 AU.

In §2, we describe our observations of comet Q2/Machholz. In §3 we compare the mixing ratios of several parent volatiles with respect to $H_2O$, measured at heliocentric distances of 1.49 AU (Nov. 2004) and 1.21 AU (Jan. 2005). In §4 we discuss our detection of $CH_4$ and simultaneous search for $CH_3D$. In §5 we estimate the $CH_4$ spin temperature, followed by discussion of results (§6).

---

[2] Hereafter Q2/Machholz or C/2004 Q2

## 2. OBSERVATION OF C/2004 Q2 (MACHHOLZ) WITH NIRSPEC AT KECK 2

Comet C/2004 Q2 (Machholz) was discovered on August 27, 2004 (Machholz et al. 2004) and reached naked-eye brightness (V < 4$^m$) by January 2005. Q2/Machholz belongs to the dynamical class of "nearly isotropic comets", whose dynamical reservoir in the present-day Solar system is most likely the Oort cloud (Levison 1996). It has a highly eccentric (e > 0.999) orbit. As a result of the comet's passage in the inner solar system, the 1/a orbital parameter changed from 0.000408 to 0.0004433. The comet's closest approach to Earth (geocentric distance $\Delta$ = 0.35 AU) occurred on Jan. 5, 2005, and its perihelion was on Jan. 24, 2004 at heliocentric distance $R_h$ = 1.205 AU (see Nakano 2006 and the JPL Small Body database [http://ssd.jpl.nasa.gov/]).

Via different methodologies three optical studies revealed the following rotation periods ($P$) for the nucleus of C/2004 Q2. Farnham et al. (2007) reported $P$ = 17.60 ± 0.05 h based on CN coma morphology. Sastri et al. (2005) reported $P$ = 9.1 ± 1.9 h based on modeling dust fans visible in R-band images. Reyniers et al. (2009) reported $P$ = 9.1 ± 0.2 h based on light curve analysis of optical broadband images. The latter study considers the possibility that their method has sampled one half instead of the full period.

We observed Q2/Machholz with NIRSPEC at the Keck-2 telescope on November 28-29, 2004 and on January 19, 2005 (UT dates). Table 1 shows condensed observing logs and derived water production rates. During the first observing run we quantified the production rates and relative abundances of several species commonly observed at IR wavelengths, including $H_2O$, $CH_4$, HCN, $C_2H_6$, $CH_3OH$, CO, and $H_2CO$. The observing circumstances were exceptionally favorable in January, allowing (in addition to measuring the overall volatile composition) detections of weaker species (acetylene, ammonia) and a sensitive search for mono-deuterated methane.

We nodded the telescope along the 24" long slit in an $A_1B_1B_2A_2$ sequence with 12" beam separation. The operation ($A_1 - B_1 - B_2 + A_2$) provided cancellation (to second order in air mass) of thermal background emission and of "sky" line emission from the Earth's atmosphere. A slit width of 0.43" resulted in spectral resolving power $\lambda/\delta\lambda \approx$ 25,000.

The cross-dispersed capability of NIRSPEC permitted sampling of all targeted molecules using two instrument settings in the L-band (2.9 – 4.0 μm) and one in the M-band (4.4 – 5.5 μm). The "KL1" setting samples (simultaneously) $H_2O$, $C_2H_6$, and $CH_3OH$; "KL2" samples $H_2O$, HCN, $NH_3$, $C_2H_2$, $C_2H_6$, $CH_4$, $CH_3D$ and $H_2CO$; and "MWA" samples $H_2O$ and CO. The ability to always simultaneously measure water within the same grading setting is significant. As the dominant parent volatile $H_2O$ serves as the natural "baseline" with which the abundances of all other species are compared. Simultaneous sampling of $H_2O$ and other species eliminates most sources of systematic error that otherwise would affect the resulting mixing ratios ($CH_4/H_2O$, $HCN/H_2O$, etc.).

Data reduction, flux calibration (based on observations of a standard star), and spectral extraction were achieved using custom-designed algorithms developed specifically for our comet observations. These algorithms are described in multiple sources, including DiSanti et al. (2006), Villanueva et al. (2009) and references therein. Bonev (2005; Appendix 2) reviews in detail all important steps leading from data acquisition to flux calibrated spectra of C/2004 Q2. Wavelength calibration is accomplished by comparing the sky radiance with spectra synthesized using a rigorous line-by-line radiative transfer model of the terrestrial atmosphere (GENLN2; Edwards 1992). This model was recently updated to properly include pressure-shift coefficients and the latest spectroscopic parameters (Villanueva et al. 2008; Hewagama et al. 2003).

Flux-calibrated spectra of C/2004 Q2 obtained on Nov. 28-29, 2004 and on Jan. 19, 2005 are shown in Figure 1[3]. The January spectra are of especially high quality allowing detection of $C_2H_2$ and $NH_3$.

3. ORGANIC VOLATILE COMPOSITION

Production rates (Q, molecules $s^{-1}$) were obtained by comparing the measured line fluxes to predicted fluorescence efficiencies (g-factors) for the appropriate rotational temperature ($T_{rot}$). The *g*-factors are based on quantum-mechanical fluorescence models for $H_2O$ (Dello Russo et al. 2004, 2005), $C_2H_6$ (Dello Russo et al. 2001), $CH_4$ (Gibb et al. 2003), $CH_3OH$ (Reuter et al. 1992; DiSanti et al. 2002), $H_2CO$ (DiSanti et al. 2006; Reuter et al. 1989), CO (DiSanti et al. 2001), HCN (Magee-Sauer et al. 1999a), $C_2H_2$ (Magee-Sauer et al. 2002), and $NH_3$ (Magee-Sauer at al. 2008).

---

[3] Selected flux-calibrated spectra in ascii format can be requested from the authors. All flux calibrated spectra will be made available after the on-going investigations on Q2/Machholz that are not included in this work are complete.

Multiple lines are used to obtain most production rates and the level of disagreement between line-by-line production rates is included in the uncertainty. The exceptions are $CH_3OH$ and $H_2CO$ whose production rates are based on the integrated intensity of the detected Q-branches.

The production rate of ammonia deserves a special discussion because of possible spectral "contamination" from other species. First, we find no evidence that the emission near 3295 $cm^{-1}$ labeled $NH_3$ (Fig. 1g) is contaminated by $NH_2$. Both the measured frequency (3295.4 $cm^{-1}$) and the line width of this spectral feature are entirely consistent with this emission originating from two blended ammonia lines [aqP(2,0) and aqP(2,1); $\nu_1$ band]. On the other hand, the $NH_2$ ($\nu_3$ band) emission should peak at 3295.5 $cm^{-1}$. Second, the contribution of HCN to the emission near 3317 $cm^{-1}$ is negligible (nevertheless the HCN contribution was modeled out before quantitative analysis of $NH_3$). This weak feature is primarily due to ammonia [sqP(1,0)]. We note that definitive searches for $NH_2$ and detection of $NH_3$ via larger number of (stronger) lines requires higher spectral resolving power and broader spectral coverage.

Detailed descriptions of the methodology for obtaining rotational temperatures, production rates, and mixing ratios (or their upper limits) are given elsewhere (Villanueva et al. 2009; DiSanti et al. 2006; Bonev et al. 2006, 2007; Dello Russo et al. 2004). Production rates, rotational temperatures, and mixing ratios (relative to $H_2O$) are summarized in Tables 2a and 2b. The observed transitions of $H_2O$, HCN, and CO sample quantum levels with a range of rotational energies sufficiently broad to constrain the temperature. We adopted the measured temperatures for water for all other species. Noting that $T_{rot}$(HCN) and $T_{rot}$($H_2O$) differ significantly in January, we verified that the mixing ratio of HCN is only weakly influenced by this difference. We sample enough lines of HCN so the resulting production rate is relatively insensitive to moderate (within 15-20 K) changes in rotational temperature. As a result, the relative abundance $HCN/H_2O$ is nearly the same whether we assume $T_{rot}(HCN) = 76$ K or $T_{rot}(HCN) = 93$ K (Table 2a).

Our water production rate derived for Jan. 19, 2005 agrees within error with the Odin satellite submillimeter measurement (Jan. 20, 2005) – Biver et al. (2007) report $Q(H_2O) = (26.4 \pm 0.8) \times 10^{28}$ s$^{-1}$. Most importantly, the agreement of our mixing ratios measured at $R_h = 1.49$ AU and 1.21 AU (Fig. 3) is consistent with chemical homogeneity in "mean volatile release" from the nucleus of C/2004 Q2 (see discussion in §6.2).

4. AN UPPER LIMIT FOR $CH_3D/CH_4$ IN COMET C/2004 Q2 (MACHHOLZ)

Figure 2 shows our detection of $CH_4$ and simultaneous search for $CH_3D$ in C/2004 Q2. A production rate for methane is derived from the KL2 setting based on the R0, R1, and R2 lines, and from the KL1 setting, based on the P2 line (see Fig. 1a). The $CH_4$ production rates are obtained independently for each NIRSPEC setting and agree within error (Table 3).

In our search for $CH_3D$, we apply the following criterion for molecular detection:

(1) More than one emission line of the searched molecule must be detected.
(2) The abundances of the searched volatile, derived independently from each detected line must be in reasonable agreement.

A $CH_3D$ fluorescence model (accounting for terrestrial atmospheric transmittance) is shown on Figure 2. This model is adopted from Kawakita & Watanabe (2003). We notice an emission feature close in frequency to the $CH_3D$ transition ($^RR(0,0)$) near 3025 cm$^{-1}$. If we assume that this emission belongs to $CH_3D$, we obtain $CH_3D/CH_4 = 0.3$ for the abundance of mono-deuterated methane. However, this value is inconsistent with the non-detections of other sampled transitions (see model Fig. 2), and therefore violates both detection criteria (1) and (2). For such a high abundance, the strong $CH_3D$ line ($^RR(3,3)$) near 3054 cm$^{-1}$ is predicted to be of comparable intensity with the lines from the OH "quadruplet" between 3047 and 3043 cm$^{-1}$ (as shown on the figure). Instead, the non-detection of this line implies that $CH_3D/CH_4 < 0.027$ (3σ limit). Although we detect a spectral feature at the frequency of one searched $CH_3D$ line, we cannot claim detection of this molecule, because other emissions that are predicted to be substantially stronger are not identified.

Three factors affect the sensitivity of the resulting upper limit for mono-deuterated methane. First is the photon noise near the expected positions of $CH_3D$ lines. This noise is dominated by thermal background emission and by the "sky" line emission from the terrestrial atmosphere (e.g., Fig. 1 of Bonev et al. 2006). The second factor is possible frequency overlap by emission from species other than $CH_3D$ in the spectrally-crowded region near 3.3 μm. Using only transitions without identified blends with emission from other species (these are marked with "X" in Fig. 2), we obtained a 3σ upper limit of $CH_3D/CH_4 < 0.020$ (Table 3), corresponding to D/H (methane) < 0.005[4].

The third factor affecting the reported upper limit is the adopted rotational temperature for $CH_3D$. We assumed the same $T_{rot}$ (93 K) for both $CH_4$ and $CH_3D$, equal to that measured for $H_2O$ (Table 2a). This represents the more conservative choice for deriving an upper limit – assuming lower rotational temperature for $CH_3D$ would result in a smaller (more stringent) upper limit for $CH_3D/CH_4$, because the g-factors of the sampled strong $CH_3D$ lines increase with decreasing $T_{rot}$ (see Kawakita & Watanabe 2003).

Kawakita & Kobayashi (2009) published a completely independent investigation focusing on the spin temperatures of water and methane in Q2/Maccholz and on $CH_3D$. They observed the comet on Jan. 30, 2005 with NIRSPEC and accumulated 36 minutes on source (vs. 8 minutes in our study). Based on a tentative detection of the $^RR(3,3)$ line, this work reports D/H = 0.0038 ± 0.0013 ($CH_3D/CH_4$ = 0.0152 ± 0.0052). These authors point out that the detected emission might be attributed to another species (e.g. $CH_3OH$), in which case they conclude that on the 95% confidence level D/H < 0.0064 ($CH_3D/CH_4 < 0.025$). These results are in good agreement with our retrieval, thereby increasing the reliability of D/H measured for methane in C/2004 Q2.

Finally, our 3σ upper limit for $CH_3D/CH_4$ (0.020) agrees with the measurement in another Oort cloud Comet - C/2001 Q2 (NEAT) - reported by Kawakita et al. (2005) ($CH_3D/CH_4 < 0.04$, 95% confidence limit).

---

[4] D/H(methane) = 0.25 x $CH_3D/CH_4$.

5. SPIN TEMPERATURE OF METHANE IN C/2004 Q2: CONSTRAINTS AND UNCERTAINTIES.

In this section we describe the constraints on the spin temperature of $CH_4$ that can be imposed by the C/2004 Q2 measured spectra. The $CH_4$ molecule exists in three types of nuclear spin species (A, E, and F; see Barnes et al. 1972). The temperature that reproduces a given A:E:F abundance ratio under conditions of thermal equilibrium is defined as the methane spin temperature (Kawakita et al. 2005, Gibb et al. 2003). This parameter should be distinguished from the rotational temperature defined as the Boltzmann temperature that describes the rotational population distribution within the ground vibrational level for a given spin "ladder". Radiative and collisional transitions among levels from different spin states are strongly forbidden.

5.1     The R0/R1 line ratio as a spin temperature diagnostic

The R0 and R1 lines ($v_3$ band) of $CH_4$ represent pure A and F transitions respectively ($A_2 - A_1$ and $F_2 - F_1$; see Gibb et al. 2003 or Drapatz et al. 1987), and their flux ratio is sensitive to $T_{spin}$. The abundance ratio F/A increases with spin temperature until reaching the statistical equilibrium value of 9:5 at $T_{spin} \approx 45$ K (see for example Fig. 4 in Gibb et al. 2003). The strong R0 and R1 lines are detected simultaneously (Fig. 2). Their flux ratio was determined with good precision and compared to the predicted flux ratios, that depend on both spin and rotational temperature (Gibb et al. 2003). We tested a range of spin temperatures, and (initially) adopted the rotational temperatures found for water measured simultaneously (Table 2a). Measured and predicted line ratios are summarized in Table 4.

We found that the R0/R1 line ratio was consistent with spin temperatures higher than 35 K for both dates (Nov. 28, 2004 and Jan. 19, 2005). Our January observations rule out $T_{spin}$ below 38 K at the 99% confidence limit imposed by the 3σ stochastic noise uncertainty (≈12%) in the measured line ratio. For November, the observed line ratio differs from the statistical equilibrium value by only 1.1σ (the 3σ stochastic noise limit is $T_{spin} > 35$ K). This result is consistent with the independently derived value by Kawakita & Kobayashi (2009), who report $T_{spin}(CH_4) > 36$ K at the 95% confidence limit, based on $\chi^2$ retrieval.

We verified that the assumed rotational temperature for methane does not impact our conclusion for $T_{spin} > \sim35$ K. This conclusion holds for a wide range of $T_{rot}$ as shown on Table 4 and Fig. 4. Assuming a rotational temperature substantially lower than that found for $H_2O$ reduces the agreement between observed and predicted line ratios for any value of the spin temperature. However, the predicted line ratio (R0/R1) for statistical equilibrium still provides the best match to the measurement. Assuming rotational temperatures higher than measured for water strongly favors $T_{spin} > \sim35$ K.

We also verified the R0/R1 ratio (and respectively $T_{spin}$) is insignificantly affected by variation in the telluric transmittance function. The reason is that a change in the telluric transmittance produces correlated change in the derived top-of-the-atmosphere fluxes of the two methane lines, so their relative intensities are only weakly affected. Finally, similar lower limits for $T_{spin}$ are found at both blue (Nov. 28, 2004) and red (Jan. 19, 2005) Doppler shifts also in favor of the measurement's reliability.

5.2     Limitations of the spin temperature retrieval.

The main limitation of the derived spin temperature are: (1) it is based on a single line of each spin species sampled (A and F for R0 and R1 respectively), and (2) a common adopted rotational temperature is assigned to each spin ladder. Most $CH_4$ lines are severely extinguished by telluric absorption at the Doppler shifts of our observations (given in Table 1). As a result only R0 and R1 are detected with high signal-to-noise, required for a sensitive $T_{spin}$ measurement.

Even if the Doppler shift was sufficient to obtain high signal-to-noise for a large number of methane lines, emissions other than R0 and R1 contain lines representing transitions from different spin ladders (Gibb et al. 2003) that remain blended at the spectral resolving power of NIRSPEC. In practice, it is more difficult to derive unique values for $T_{rot}(CH_4)$ and $T_{spin}(CH_4)$ from such blended transitions. Robust determination of $T_{rot}(CH_4)$ and $T_{spin}(CH_4)$ requires detections of multiple lines of spectrally-resolved A-, F-, and E-type transitions that sample quantum states having a broad range of rotational energies. This would allow testing whether the rotational population distributions within a given spin ladder can be characterized by a single rotational temperature and whether the $T_{rot}$ measured independently for the three spin ladders agree. Only then can $T_{spin}$ be determined more reliably.

Our retrieval does not allow a robust line-by-line test of the $CH_4$ fluorescence model at hand, similarly to the way it has been done for $H_2O$ (e.g. Dello Russo et al. 2004; Bonev et al. 2007). For this reason the reported error in $T_{spin}(CH_4)$ does not include a potential model-related uncertainty.

We defer more detailed discussion on problems related to $CH_4$ spin temperature analysis in comets to a separate paper (Gibb, E. L. et al. 2009, in preparation). Although potentially less robust, the value derived here for $T_{spin}(CH_4)$ is in very good agreement with the result derived for $H_2O$ in C/2004 Q2. Bonev et al. (2007) report $T_{spin}(H_2O) > 34$ K (this limit is dominated by the systematic uncertainty in line-by-line analysis which exceeds the uncertainty due to photon noise), while Kawakita & Kobayashi (2009) report $T_{spin}(H_2O) > 27$ K (95 % confidence limit, based on $\chi^2$ retrieval).

6. DISCUSSION

6.1 Comparison with organic volatile abundances measured in other comets.

The abundances of $C_2H_6$, $CH_3OH$, and HCN relative to $H_2O$ in C/2004 Q2 are similar to those observed in several Oort cloud comets, tentatively referred to as "organics-normal" (see Mumma et al. 2008, DiSanti & Mumma 2008, Mumma et al. 2003). The abundance of native $H_2CO$ is at the lower end of the range observed in Oort cloud comets (~0.1% – ~0.8%).

The mixing ratios $CO/H_2O$ and $CH_4/H_2O$ were found to vary by an order of magnitude (or more in the case of CO) among comets (Mumma et al. 2003; Gibb et al. 2003, 2007b). However, there is no correlation is between the CO and $CH_4$ abundances. This notion is supported by our observations of Q2/Maccholz where the mixing ratio $CO/H_2O$ ($\approx 5$ %) is intermediate in comparison with other comets (abundances vary between < 0.1% to ~15%), while $CH_4/H_2O$ (~1.45%) is near the high end of the observed range (< ~0.1% – ~2.0 %).

Interestingly, acetylene is low in abundance relative to both water and ethane. The $C_2H_6/C_2H_2$ ratio (~5) in Q2/Machholz is substantially larger than observed in other comets. The relative abundance $CH_4/HCN$ (~10) is also distinctly larger than in previous comets observed (~2.5 to ~5.5).

More detailed inter-comparison among comets observed in the IR (including C/2004 Q2) will be presented in a future review dedicated to the emerging taxonomic classification of comets based on parent volatile composition (see Mumma et al. 2008 for preliminary results). Here, we focus our discussion on two particular topics: evidence for compositional homogeneity in Q2/Machholz (§6.2), and cosmogonic implications of our measured $CH_3D/CH_4$ upper limit (§6.3-6.5).

6.2     Compositional homogeneity of mean volatile release of C/2004 Q2

Infrared measurements of parent volatiles are rarely obtained over a large range of heliocentric distances for a given comet (see §3), owing to the difficulty in scheduling adequate observing time. Comets 1P/Halley and C/1995 O1 (Hale-Bopp) are exceptions. Both featured sufficient advance notice to permit planned campaigns. For Halley, water was studied from 1.16 AU (pre-perihelion), through perihelion (0.58 AU), and out again to 1.10 AU (post-perihelion). The ortho-para ratio of water remained unchanged over that interval (Mumma et al. 1993).

In Hale-Bopp, CO and $C_2H_6$ were measured from respectively 4.1 AU and 3.01 AU pre-perihelion, through perihelion (0.91 AU), and out again to 2.83 AU post-perihelion (DiSanti et al. 2001, Dello Russo et al. 2001). The specific mass loss over this period (kg/sec * $R_h^2$, with $R_h$ in AU) was discussed and compared with similar measures of dust and other parent volatiles (Mumma et al. 2003). For comparison, in the radio Biver et al. (1999) studied extensively the long-term evolution of outgassing in Hale-Bopp between 7.0 AU (pre-perihelion) to 4.0 AU (post-perihelion). These authors investigated the evolution of production rates and relative abundances of a number of species, founding that the latter did not exhibit substantial changes at $R_h$ < 1.5 AU. Such studies address (at least) two important questions:

1. Is the comet nucleus compositionally homogeneous (or heterogeneous) on a bulk scale?
2. Are the observed chemical abundances of comet volatiles affected by heliocentric evolution of outgassing?

Several other comets were sampled at IR wavelengths at diverse heliocentric distances. In Comet C/2001 A2 (LINEAR), the $H_2CO/H_2O$ ratio varied day-to-day by a factor of four at 1.16 AU (while the comet was in outburst), while the $CH_4/H_2O$ ratio increased by a factor of two as the comet moved from ~1.16 AU and 1.55 AU (Gibb et al. 2007b). In 9P/Tempel-1, "Deep Impact" returned evidence of dissimilar $CO_2/H_2O$ ratios in individual vents (Feaga et al. 2007). Such variability might be attributed to internal heterogeneity in volatile composition caused by radial dynamical mixing of cometesimals prior to forming a cometary nucleus.

By contrast, Comet 8P/Tuttle was observed over a range of heliocentric distances (1.16 – 1.08 AU, spanning 38 days) during its favorable 2007-2008 apparition and very similar relative abundances resulted from all IR observations (Bonev et al. 2008, Böhnhardt et al. 2008), consistent with homogeneous volatile composition.

The split comet 73P/Schwassmann-Wachmann 3 (73P/SW3) provided the strongest evidence for compositional homogeneity. The mixing ratios of multiple parent volatiles measured in fragments B and C were remarkably similar (heliocentric distances 1.06 to 0.99 AU, Dello Russo et al. 2007). The mixing ratios $C_2H_6/H_2O$ and $HCN/H_2O$ measured independently in fragment C (Villanueva et al. 2006, based on measurements at heliocentric distances 1.27 and 1.19 AU) and in fragment B (heliocentric distance 1.03 AU, Kobayashi et al. 2007) agree with one another and with those reported by Dello Russo et al. thereby increasing the evidence for compositional homogeneity of the parent body on bulk scales.

The mixing ratios of parent volatiles in C/2004 Q2 measured at heliocentric distance of 1.5 AU agree with those measured at 1.2 AU (Figure 3), although separated in time by nearly two months. During this interval the comet's gas productivity increased by a factor of about 2 (Tables 1, 2), while the abundances relative to $H_2O$ remained (nearly) unchanged. The similarity in relative abundances measured at 1.5 AU and 1.2 AU contrasts strongly with the degree of chemical diversity observed among Oort cloud comets (see §1).

Our results support chemical homogeneity in the *mean volatile release* from C/2004 Q2, but do not exclude the possibility that the nucleus has more than one active region (perhaps many) as suggested by studies of daughter fragments (Farnham et al. 2007; Lin et al. 2007). Even if different active regions are characterized by distinct mixing ratios among volatiles, the "mean" release exhibits nearly identical chemistry suggesting that roughly the same proportions of release are maintained. Such compositional similarities for a given comet strongly suggest that the heliocentric evolution of outgassing [whether gradual (e.g., 8P/Tuttle and Q2/Machholz) or abrupt (from outburst or splitting as in 73P/SW3)] has not affected the observed abundances. The most plausible hypothesis is that these abundances are representative of the bulk volatile composition and reflect the early history of cometary ices.

6.3     The D/H ratio in methane as a cosmogonic "thermometer"

Searches for deuterated species are a key part in testing the "interstellar-comet connection" (Charnley and Rodgers 2008a,b). In particular:

1. What is the processing history experienced by interstellar organic matter as it is incorporated into the disk of a (low mass) protostar?
2. How (and to what extent) is the deuterium fraction modified (with respect insterstellar values) prior to incorporation of ices in cometary nuclei?
3. Does "unprocessed" (without change in isotopic signatures) interstellar volatile material exist in comets?

Models that incorporate gas phase chemistry and trace the evolution of D/H from cold (10 K) molecular cloud, through core collapse, to formation of proto-planetary disk predict $CH_3D/CH_4 \geq 0.10$ (Aikawa & Herbst 1999, 2001; see also the nice discussion in Kawakita et al. 2005). On the other hand, our measured upper limit ($CH_3D/CH_4 < 0.02$, $3\sigma$) agrees with predictions for methane formation via ion-molecule reactions in relatively warm (> 25-30 K) gas (Millar et al. 1989; Charnley and Rodgers 2008a). Models of gradients in temperature and composition in proto-stellar cloud cores indeed predict low D/H for these outer regions – thought to be the main source for material in the disk at the time when comets form (Charnley & Rodgers 2008b)[5]. Thus the measured upper limit for $CH_3D/CH_4$ is consistent with, but (quite possibly) does not unambiguously imply formation of cometary methane exclusively via gas phase chemistry at temperatures exceeding 25-30 K.

An important question is whether our measurement can also be explained within the interpretation of Boogert et al. (2004) who detected both gas phase and solid phase $CH_4$ along the line of sight of the massive protostar NGC 7838 IRC 9 and suggested that methane is formed by H-atom addition reactions on grain surfaces and is intimately mixed with water in the proto-stellar envelope.

Gas-grain chemistry and the effects of three desorption mechanisms (thermal-, photo-, and cosmic ray induced desorption) were incorporated by Willacy (2007) who modeled the chemical evolution in a static (i.e. w/o mixing processes, see §6.4) T-Tauri disk formed after the collapse of a molecular cloud. This model predicts $CH_3D/CH_4$ ratios in the range ~0.07 - 0.20. The inconsistency with our upper limit might be related to the initial condition ($CH_3D/CH_4 \approx 0.07$), resulting from methane formation in molecular cloud of temperature equal to 10 K.

These comparisons with chemical models suggest that the methane ice released from the nucleus of Q2/Machholz is not dominated by a component that keeps a chemical "memory" of cold (<< 25 K) environments in the natal cloud or in the outer regions of the protoplanetary disk (where similar chemistry can occur despite higher densities, see Aikawa & Herbst 2001). The $(D/H)_{H2O}$ ratio (< 2.3 x $10^{-4}$, preliminary result) in C/2004 Q2 (Biver et al. 2005) supports a similar conclusion for water. The spin temperatures of $H_2O$ (> 34 K, Bonev et al. 2007) and $CH_4$ (> 35-38 K, this work) in C/2004 Q2 also support a relatively warm formation environment, providing $T_{spin}$ is indeed a measure of the chemical formation temperature of the corresponding molecule.

---

[5] In the same model, the region characterized by highest D/H ratios is in the center of the pre-stellar cloud. Most of this material would likely be absorbed by the forming proto-star.

These conclusions are tentative because the current models for evolution of deuterium enrichments in comet volatiles are still under development. One of the principal challenges is to investigate the possibility for nebular (vs. interstellar) origin of volatile matter incorporated into comet nuclei.

6.4  The possibility for synthesis of hydrocarbons in the inner protoplanetary disk and outward radial transport of matter

An alternative pathway that can explain our $CH_3D/CH_4$ upper limit involves hydrocarbons that formed in the inner nebula. Based on laboratory simulations, Nuth et al. (2000) predicted that hydrogenated species can be synthesized along with crystalline dust in the hot environments of the inner solar nebula. Recent laboratory work (Nuth et al. 2008) demonstrated that an organics-rich macromolecular coating forms on various grain surfaces via reactions that reduce CO to produce hydrocarbons. In this process molecules, like $CH_4$ and $C_2H_6$ are released in the gas phase, while forming layers of carbon-rich macromolecular residue further providing a catalytic surface for continuing efficient synthesis of organic material. A similar process could produce hydrocarbons efficiently in the hot inner proto-planetary disk.

Outward radial transport of matter could bring these products of high-temperature chemistry to environments where ices can form. Several models (Dubrille 1993, Drouart et al. 1999, Shu et al. 2001, Bockelee-Morvan et al. 2002, Gail 2002) explored mechanisms for radial mixing in the nebula, and predicted "grand-scale" transport of dust and gas within the nebular disk allowing high-temperature chemical products to be ultimately incorporated into comet nuclei. These predictions were resoundingly confirmed by analysis of refractory minerals in 81P/Wild 2 returned by the "Stardust" mission (Brownlee et al. 2006, Zolensky et al. 2006).

We expect that material synthesized in the hot inner nebula would be characterized by the protosolar value of D/H. Our observed upper limit for $CH_3D/CH_4$ corresponds to D/H ($< 5 \times 10^{-3}$, $3\sigma$) exceeding the protosolar value ($[2.35 \pm 0.3] \times 10^{-5}$)[6] by two orders of magnitude. Therefore our result does not rule out the hypothesis that (some) methane (and other organics) in C/2004 Q2 was synthesized in the hot inner nebula. Under this scenario, the amount of inner-nebula organics stored in comet nuclei would be controlled by:

1.  the time-dependent abundances of various hydrocarbons resulting from inner-nebula chemistry,
2.  the efficiency and distance scale (neither are well-understood at present) of outward radial transport in the protoplanetary disk at a particular time, and
3.  the conditions in outer ice-forming regions, where comet nuclei could accrete.

Since crystalline dust is also produced in the inner nebula (see review of Wooden 2008), the ratio of crystalline-to-total dust content in C/2004 Q2 would provide an additional constraint on the hypothesis of nebular origin for a fraction of organic material stored in the nucleus of this comet (Drouart et al. 1999). A high fraction of crystalline dust is predicted to correlate with low CO-to-hydrocarbons (e.g $CO/C_2H_6$) and $N_2/NH_3$ ratios (Nuth et al. 2000).

6.5     Summary on constraints from $CH_3D/CH_4$

Our upper limit for $CH_3D/CH_4$ ($< 0.020$, $3\sigma$) implies that methane ice released from the nucleus of Q2/Machholz is not dominated by an unprocessed component formed in extremely cold (~10 K) environments either in the inner portions of the proto-solar molecular cloud or (at a later evolutionary stage) in the outer regions of the protoplanetary disk. We discussed three formation pathways for methane stored in the nucleus of Q2/Machholz:

1.  Formation of methane in a relatively warm (> 25 K) gas via ion-molecule chemistry (consistent with the observed $CH_3D/CH_4$ upper limit).
2.  Formation of methane simultaneously with $H_2O$ via H-atom addition reactions on interstellar icy grains in the proto-solar envelope.

---

[6] Weighted mean of the values reported by Geiss & Gloeckler 1998, Mahaffy et al. 1998, and Lellouch et al. 2001; see Mousis et al. 2002.

3. Synthesis of methane and other hydrocarbons in the hot inner solar nebular, followed by outward radial transport of matter (characterized by proto-solar D/H) to environments where comet nuclei can form.

Evaluating the relative contributions of these pathways and the extent to which each pathway is consistent with both the measured upper limit for $CH_3D/CH_4$ and the observed volatile abundances relative to $H_2O$ requires further modeling of chemistry including both gas-phase and gas-grain processes in the natal interstellar cloud, during core collapse, and in the protoplanetary disk.

Acknowledgements

We are grateful to Neil Dello Russo for his thorough reviews that improved the quality of the paper. We thank Hideyo Kawakita for providing an updated $CH_3D$ model. We are grateful to the following colleagues for stimulating discussions on various aspects in this work: Yuri Aikawa, Dennis Bodewits, Steven Charnley, Martin Cordiner, Joe Nuth, Diane Wooden, and Charles Woodward.

This research was supported by the NASA Planetary Astronomy, Planetary Atmospheres, and Astrobiology programs, and by the NSF Planetary Astronomy Program. The data presented herein were obtained at the W. M. Keck Observatory, operated as a scientific partnership among CalTech, UCLA, and NASA. This Observatory was made possible by the generous financial support of the W. M. Keck Foundation. The authors wish to recognize and acknowledge the very significant cultural role and reverence that the summit of Mauna Kea has always had within the indigenous Hawaiian community. We are most fortunate to have the opportunity to conduct observations from this mountain.

REFERENCES:

Aikawa, Y., & Herbst, E. 1999, ApJ, 526, 314

Barnes, W. L., Susskind, J., Hunt, R. H., & Plyler, E. K. 1972, J. Chem. Phys., 56, 5160

Biver, N., Bockelée-Morvan, D., Crovisier, J., et al. 2007, Planet. Space Sci., 55, 1058

Biver, N., Bockelée-Morvan, D., Boissier, J, et al. 2005, in Asteroids, Comets, and Meteors, *IAU Symp*. 229, abstract book, 43.

Biver, N., Bockelée-Morvan, D., Crovisier, J., et al. 2002, Earth Moon Planets, 90, 323

Biver, N., Bockelée-Morvan, D., Colom, P. et al. 1997, Earth Moon Planets, 78, 5

Bockelée-Morvan, D., Crovisier, J., Mumma, M. J., & Weaver, H. A. 2004, in Comets II, ed. M. C. Festou, H. U. Keller, & H. A. Weaver (Tucson: Univ. Arizona Press), 391

Bockelée-Morvan, D., Gautier, D., Hersant, F., Huré, J.-M., & Robert, F 2002, A&A, 384, 1107

Böhnhardt, H., Mumma, M. J., Villanueva, G. L. et al., 2008, ApJ, 683, L71

Bonev, B. P. 2005, Ph.D. thesis, Univ. Toledo, http://astrobiology.gsfc.nasa.gov/Bonev_thesis.pdf

Bonev, B. P., Mumma, M. J., Radeva, Y. L., DiSanti, M. A., Gibb, E. L., & Villanueva, G. L. 2008, ApJ, 680, L61

Bonev, B. P., Mumma, M. J., Villanueva, G. L. et al., 2007, ApJ, 661, L97

Bonev, B. P., Mumma, M. J., DiSanti, M. A. et al., 2006, ApJ, 653, 774

Boogert, A. C. A., Blake, G. A., & Öberg, K. 2004, ApJ, 615, 344

Brown, P. D., & Millar, T. J. 1989, MNRAS, 237, 661

Brownlee, D., Tsou, P., Aléon, J. et al. 2006, Science, 314, 1711

Carr J. S., & Najita, J. R. 2008, Science, 319, 1504

Charnley, S. B., & Rodgers, S. D. 2008a, Space Sci. Rev., 138, 59

Charnley, S. B., & Rodgers, S. D. 2008b, Advances in Geosciences, in press (arXiv:0806.3103)

Crovisier, J. 2006, Mol. Phys., 104, 2737

Dello Russo, N., Vervack, R. J. Jr., Weaver, H. A., et al. 2008, ApJ, 680, 793

Dello Russo, N., Vervack, R. J., Weaver, H. A. et al. 2007, Nature, 448, 172

Dello Russo, N., Bonev, B. P, DiSanti, M. A., et al. 2005, ApJ, 621, 537

Dello Russo, N., DiSanti, M. A., Magee-Sauer, K., et al. 2004, Icarus, 168, 186

Dello Russo, N., Mumma, M. J., DiSanti, M. A., Magee-Sauer, K., & Novak, R. 2001, Icarus, 153, 162

Delsemme, A. H. 2000, Icarus, 146, 313

Delsemme, A. H. 1998, Plan. Space Sci., 47, 125

DiSanti, M. A., & Mumma, M. J. 2008, Space Sci. Rev., 138, 127

DiSanti, M. A., Bonev, B. P., Magee-Sauer, K., et al. 2006, ApJ, 650, 470

DiSanti, M. A, Dello Russo, N., Magee-Sauer, K., Gibb, E. L., Reuter, D. C., & Mumma, M. J. 2002, In: Asteroids, Comets, Meteors 2002, Berlin, Germany (ESA-SP-500, p. 571)

DiSanti, M. A., Mumma, M. J., Dello Russo, N., & Magee-Sauer, K. 2001, Icarus, 153, 361

Drapatz, S., Larson, H. P., & Davis, D. S. 1987, A&A, 187, 497

Drouart, A., Dubrulle, B., Gautier, D., & Robert, F. 1999, Icarus, 140, 129

Dubrulle, B. 1993, Icarus, 106, 59

Edwards, D. P. 1992, NCAR, Technical Note 367-STR, Boulder, CO

Ehrenfreund, P., Rasmusen, S., Cleaves, J., & Chen, L. 2006, Astrobiology, 6, 490

Feaga, L. M., A'Hearn, M. F., Sunshine, J. M., Groussin, O., & Farnham, T. L. 2007, Icarus, 190, 245

Gail, H.-P. 2002, A&A, 390, 253

Gibb, E. L., Van Brunt, K. A., Brittain, S. D., & Rettig, T. W. 2007a, ApJ, 660, 1572

Gibb, E. L., DiSanti, M. A., Magee-Sauer, K., Dello Russo, N., Bonev, B. P., & Mumma, M. J. 2007b, 188, 224

Gibb, E. L., Mumma, M. J., Dello Russo, N., DiSanti, M. A., & Magee-Sauer, K. 2003, Icarus, 165, 391

Geiss, J., & Gloeckler, G., 1998, Space Sci. Rev., 84, 239

Hewagama, T. et al. 2003, Proc. SPIE, 4860, 381

Irvine, W. M., Schloerb, F. P., Crovisier, J., Fegley, Br., & Mumma, M. J. 2000, In Protostars and Planets IV, ed. V. Mannings, A. P. Boss, & S. S. Russel (Tuscon: The University of Arizona Press), 1159

Kawakita, H., & Kobayashi, H. 2009, ApJ, 693, 388

Kawakita, H., & Watanabe, J-i. 2003, 582, 534

Kawakita, H., Watanabe, J-i., Furusho, R., Fuse, T., & Boice, D. C. 2005, ApJ. 623, L49

Kobayashi, H., Kawakita, H., Mumma, M. J., Bonev, B. P., Watanabe, J-i, & Fuse, T. 2007, ApJ, 668, L75

Lellouch, E., Bézard, B., Fouchet, T., Feuchtgruber, H., Encrenaz, T., & de Graauw, T. 2001, A&A, 370, 610

Levison, H. F. 1996, in Completing the inventory of the Solar system (ASP Conference Series), 107, 173

Machholz, D. E., Garradd, G., & McNaught, R. H. 2004, IAU Circ., 8394, 1

Magee-Sauer, K., Mumma, M. J., Bonev, B. P. et al. 2008, In: Asteroids, Comets, Meteors 2008 held July 14-18, 2008 in Baltimore, Maryland. LPI Contribution No. 1405, paper id. 8054.

Magee-Sauer, K., Mumma, M. J., DiSanti, et al. 2008, Icarus, 194, 347

Magee-Sauer, K., Dello Russo, N., DiSanti, M. A, Gibb, E. L., & Mumma, M. J. 2002, In: Asteroids, Comets, Meteors 2002, Berlin, Germany (ESA-SP-500, p. 549)

Magee-Sauer, K., Mumma, M. J., DiSanti, M. A., Dello Russo, N., & Rettig, T. W. 1999, Icarus, 149, 498

Mahaffy, P. R., Donahue, T. M., Atreya, S. K., Owen, T. C., & Niemann, H. B. 1998, Space Sci. Rev., 84, 251

McLean, I. S., et al. 1998, Proc. SPIE, 3354, 566

Millar, T. J.. Bennett, A., & Herbst, E. 1989, ApJ, 340, 906

Mousis, O., Gautier, D., & Coustenis, A. 2002, Icarus, 159, 156

Mumma, M. J., Disanti, M. A., Bonev, B. P., Villanueva, G. L., Magee-Sauer, K.; Gibb, E. L. 2008, Asteroids, Comets, Meteors 2008 (Baltimore, Maryland, USA), LPI Contribution No. 1405, paper id. 8282.

Mumma, M. J., DiSanti, M. A., Magee-Sauer, K. et al. 2005, Science, 310, 270

Mumma, M. J., DiSanti, M. A., Dello Russo, N., Magee-Sauer, K., Gibb, E. L., & Novak, R. 2003, Adv. Space Res., 31, 2563

Mumma, M. J., Dello Russo, N., DiSanti, M. A., et al. 2001, Science, 292, 1334

Mumma, M. J., Weissman, P. R., & S. A. Stern 1993, in Protostars and Planets III, ed. E. H. Levy and J. I. Lunine (Univ. Ariz. Press, Tucson), 1177

Nakano, S. 2006, Oriental Astronomical Association, Nakano Notes Circular NK 1352, 20 July 2006, http://www.oaa.gr.jp/~oaacs/nk/nk1352.htm

Nuth, J. A., Johnson, N. M., & Manning, S. 2008, ApJ, 673, L225

Nuth, J. A., Hill, H. G. M, & Kletetschka 2000, Nature, 406, 275

Reuter, D. C., et al. 1989, Icarus, 95, 329

Reuter, D. C., et al. 1989, ApJ, 341, 1045

Reyniers, M., Degroote, P., Bodewits, D., Cuypers, J., & Waelkens, C. 2009, A&A, 494, 379

Sastri, J. H., Vasundhara, R., Kuppuswamy, K., Velu, C. 2005, IAU Circ., 8480, 3

Shu, F. H., Shang, H., Gounelle, M., Glassgold, A. E., & Lee, T. 2001, ApJ, 548, 1029

Villanueva, G. L., Mumma, M. J., Bonev, B. P. et al. 2009, ApJ, 690, L5

Villanueva, G. L., Mumma, M. J., Novak, R. E., & Hewagama, T. 2008, Icarus, 195, 34

Villanueva, G. L., Bonev, B. P., Mumma, M. J. et al. 2006, ApJ, 650, L87

Willacy, K. 2007, ApJ, 660, 441

Wooden, D. H. 2008, Space Sci. Rev., 138, 75

Zolensky, M. F., Zega, T. J., Yano, H. et al. 2006, Science, 314, 1735

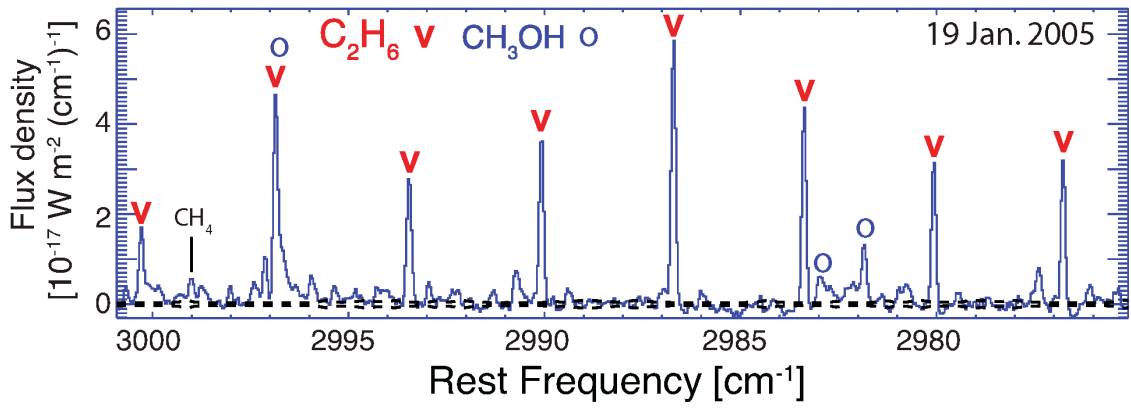

Figure 1a

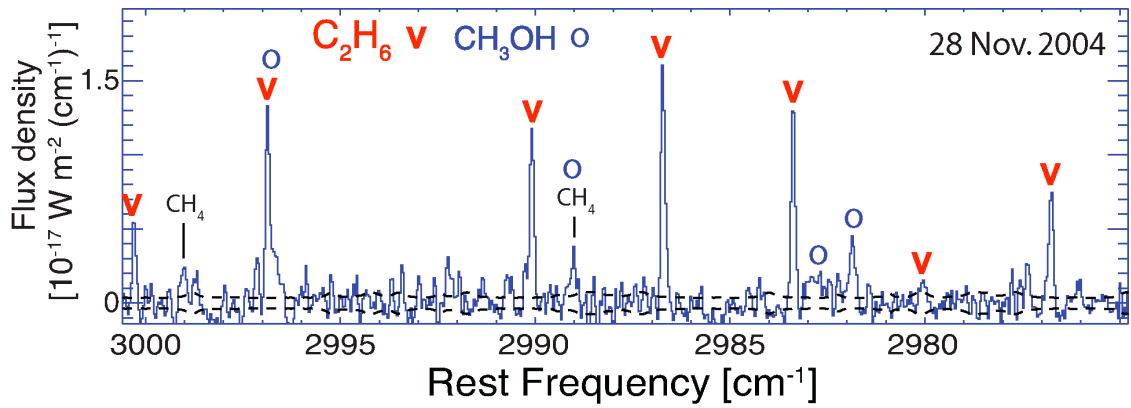

Figure 1b

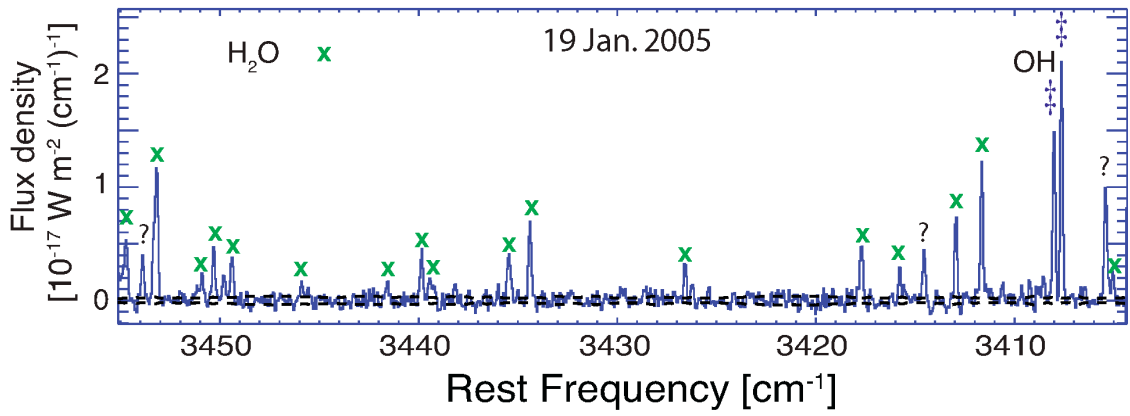

Figure 1c

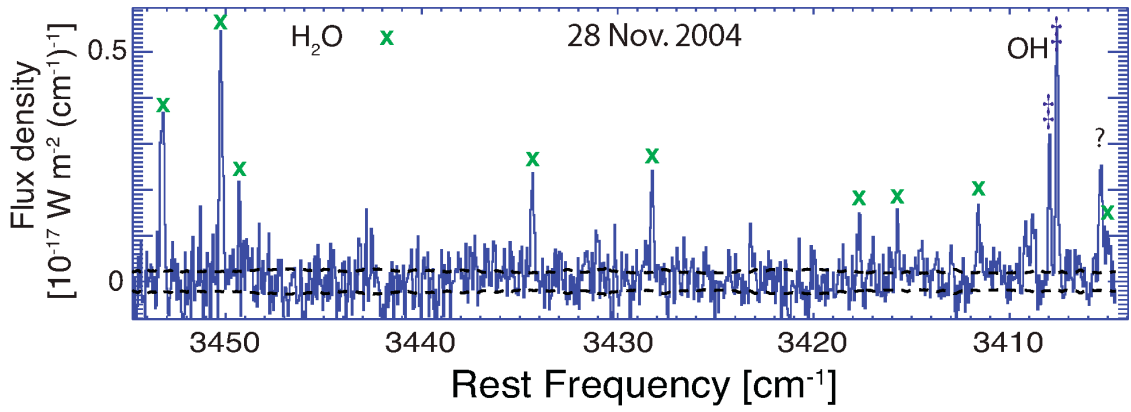

Figure 1d

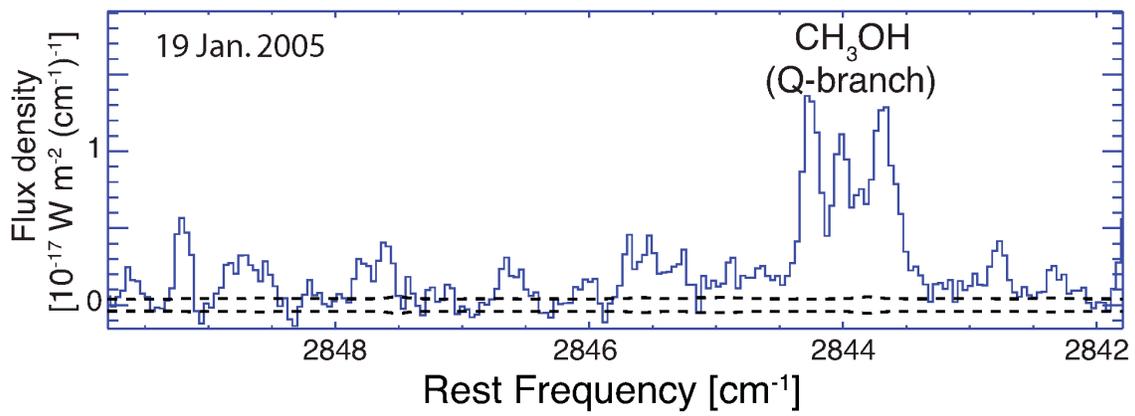

Figure 1e

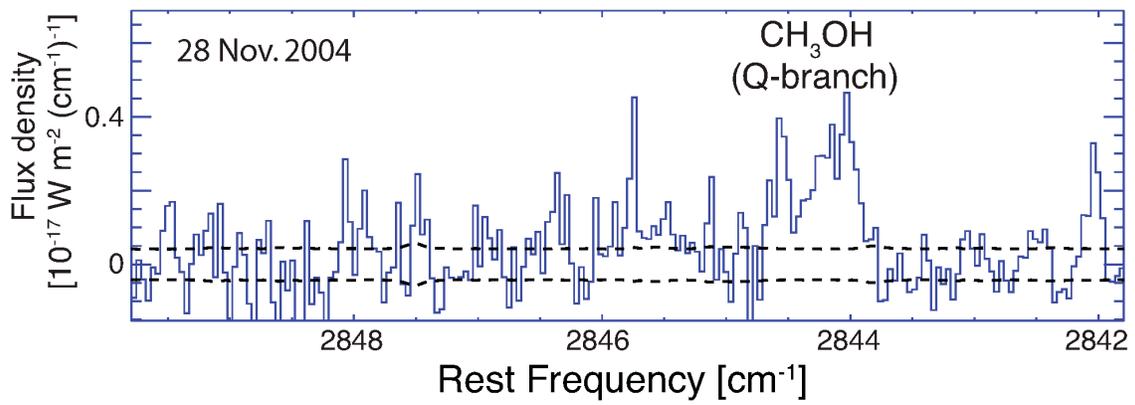

Figure 1f

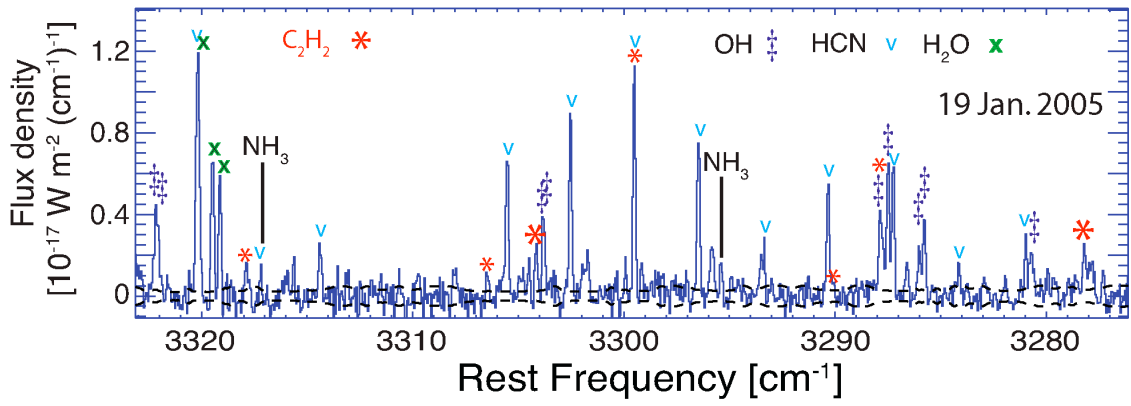

Figure 1g

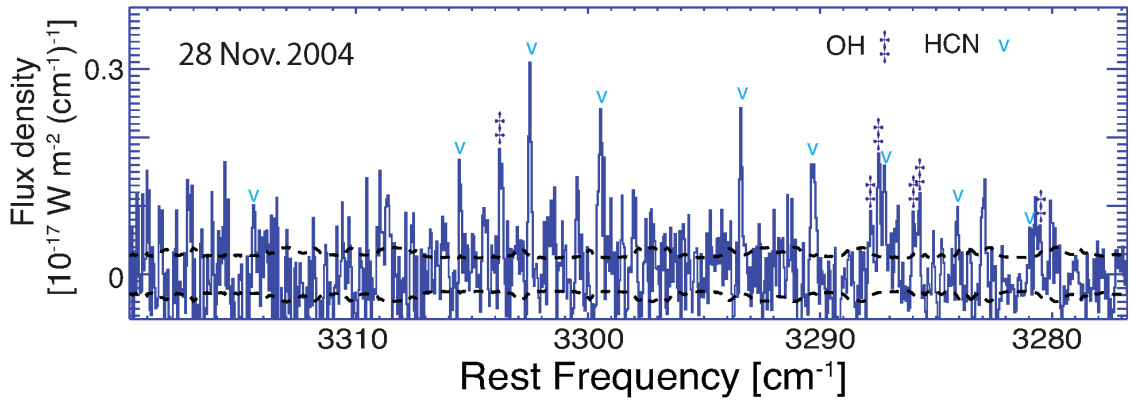

Figure 1h

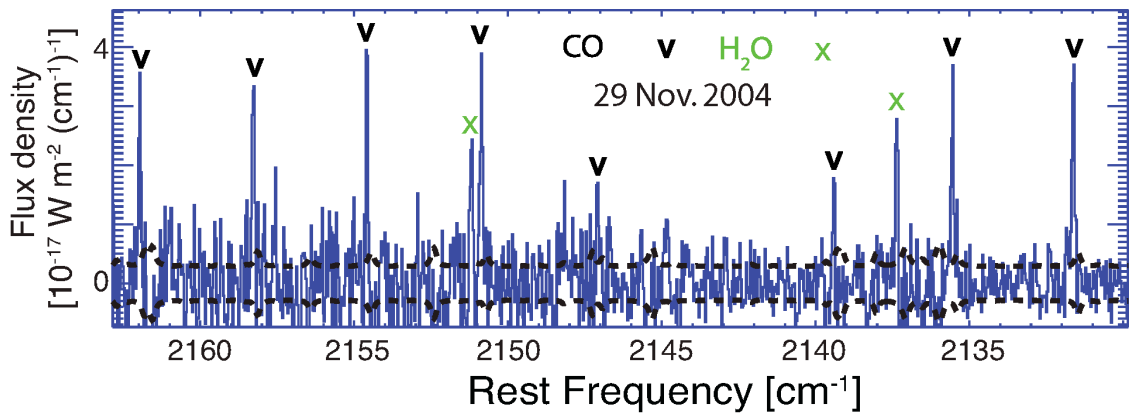

Figure 1i

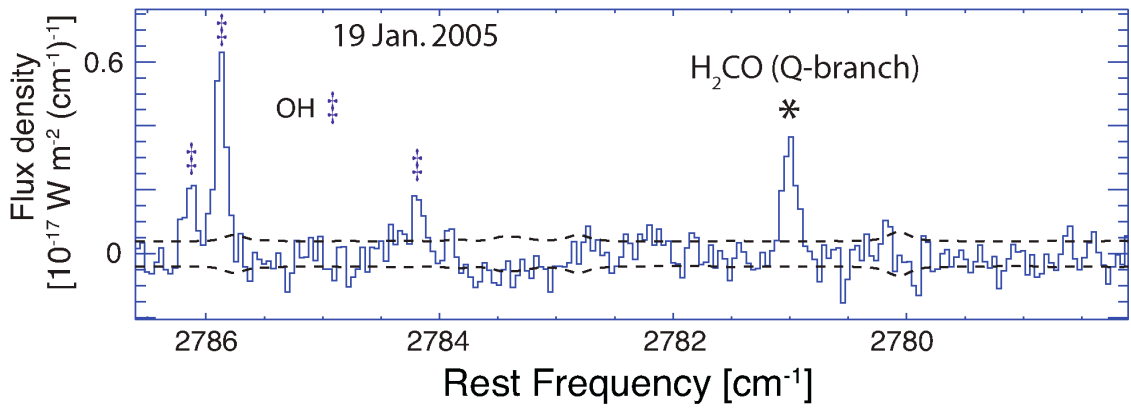

Figure 1j

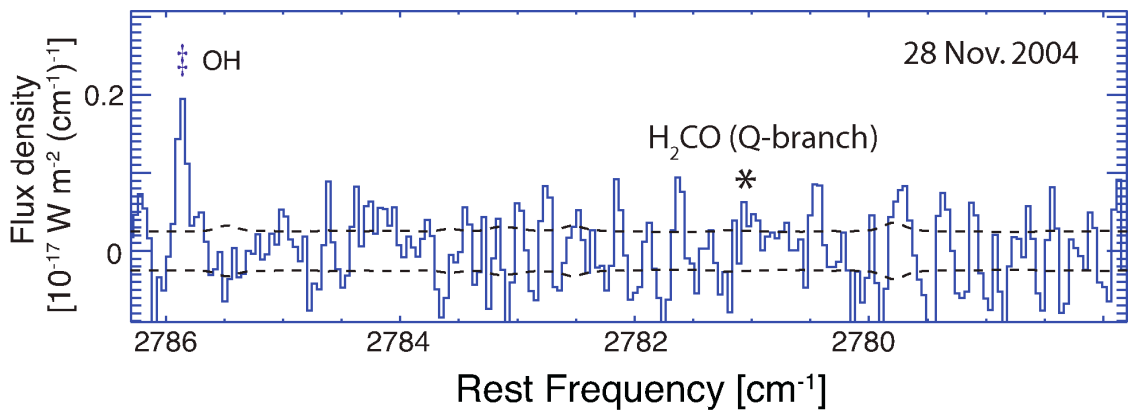

Figure 1k

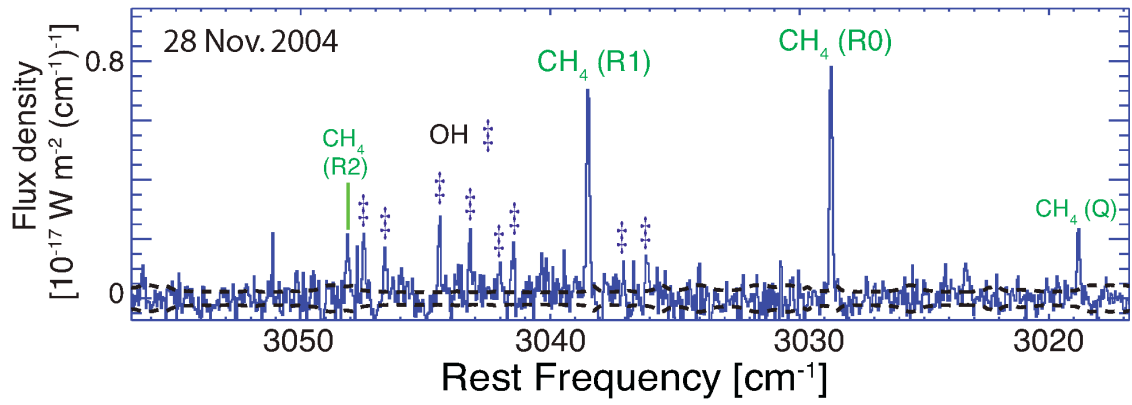

Figure 11

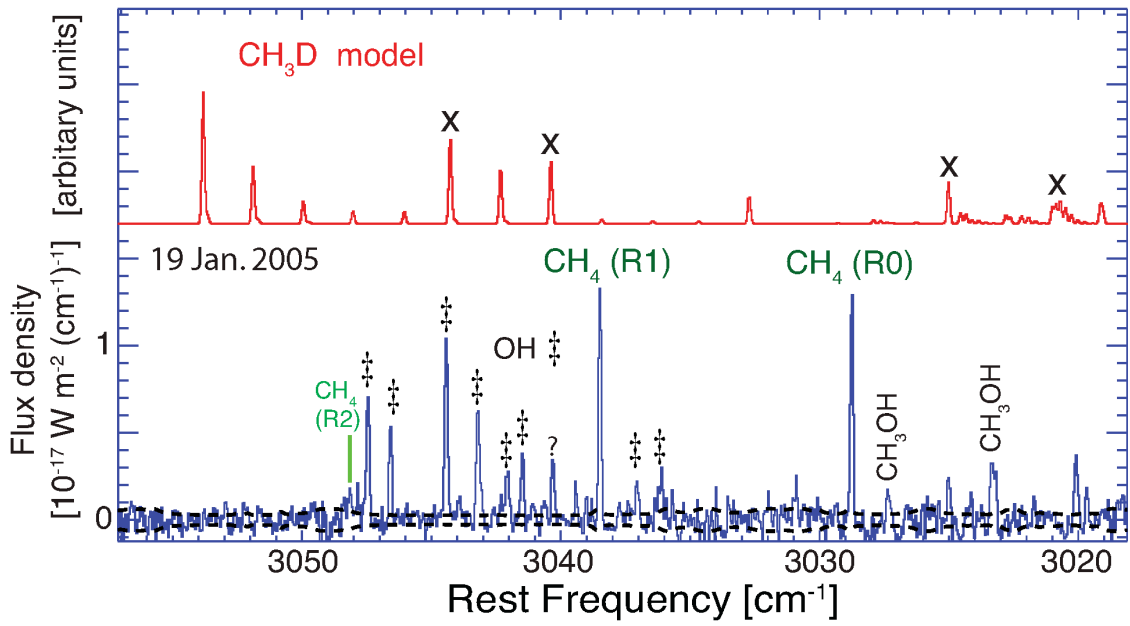

Figure 2

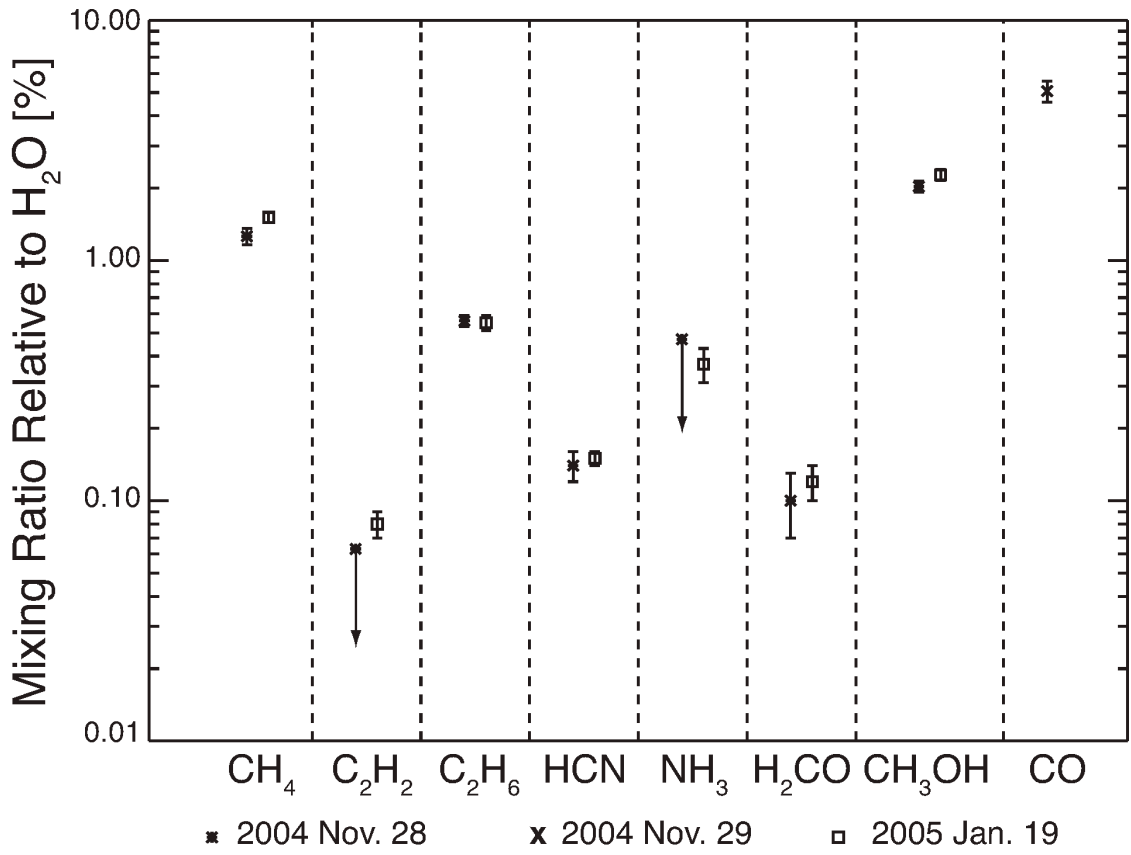

Figure 3

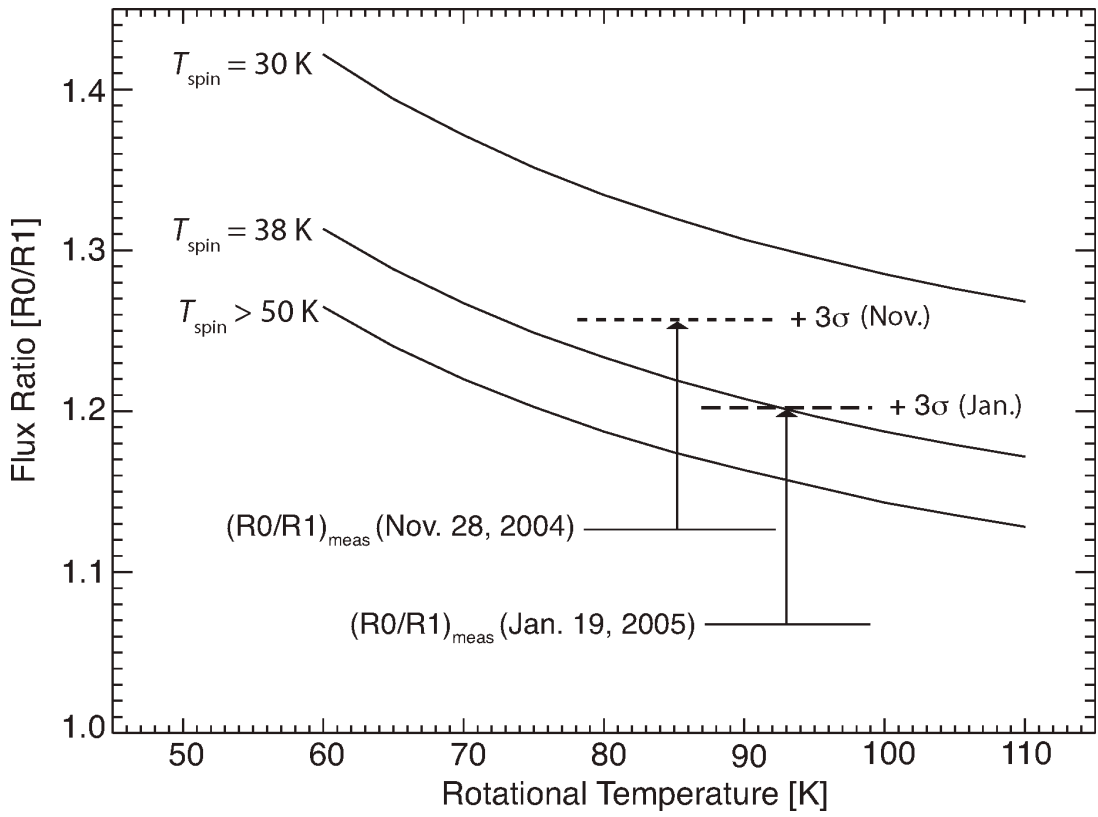

Figure 4

FIGURE CAPTIONS:

FIG.1. – The spectral gallery of C/2004 Q2 (Machholz), showing the (continuum-subtracted) signal within a 0.43x1.78 arc-second aperture centered on the comet (see for example Bonev et al. 2006, Villanueva et al. 2008). The black dashed lines envelope the photon noise (± 1σ). Detections of $C_2H_6$ (a-b), $H_2O$ (c-d), $CH_3OH$ (e-f), HCN (g-h), CO (i), $H_2CO$ (j-k), and $CH_4$ (l) are highlighted. The Q-branch of the $CH_3OH$ $\nu_3$ band (panels e-f) is commonly used to measure methanol production rates in comets from infrared observations. Acetylene (ν3 band) and ammonia (ν1 band) are detected on Jan. 19, 2005 (panel g). $C_2H_2$ (ν2 + ν4 + ν5 band) has a line at the frequency of the emission feature near 3295.8 cm$^{-1}$. However claiming detection requires developing a fluorescence model and testing it against multiple lines, so this feature is not used to derive a production rate for acetylene. The emission near 3302 cm$^{-1}$ might be a signature of $NH_2$ (ν1 band). However, $C_2H_2$ (R2, ν3 band) contributes substantially to the detected emission, which complicates quantitative analysis.

FIG.2. – Detection of $CH_4$ and search for $CH_3D$ in C/2004 Q2 on 19 Jan. 2005. The lines in the model marked with "x" are excluded from quantitative analysis because of possible spectral contamination by emissions from other species (see text for detailed discussion). The black dashed lines envelope the photon noise (± 1σ).

FIG.3. – Mixing ratios (relative to $H_2O$) measured in C/2004 Q2 (Machholz). All parent molecules except CO were measured both in November 2004 ($R_h$ = 1.49 AU) and January 2005 ($R_h$ = 1.21 AU). See discussion in §6.1 and §6.2.

FIG.4. – Dependence of the methane R0/R1 flux ratio ($CH_4$, $\nu_3$ band) on rotational and spin temperatures. Theoretical curves (following Gibb et al. 2003) for $T_{spin}$ = 30 K, $T_{spin}$ = 38 K, and $T_{spin}$ > ~ 50 K (statistical equilibrium) are shown. The measured flux ratios (R0/R1)$_{meas}$ for Nov. 28, 2004 and Jan. 19, 2005 and their +3σ stochastic uncertainties are indicated with solid and dashed horizontal lines respectively. This graphic illustrates that the observed $CH_4$ R0/R1 flux ratios are consistent with $T_{spin}$ >

35-38 K for a broad range of rotational temperature, including the rotational temperatures derived on those dates for water. See §5.2 for discussion on the possible limitations of the spin temperature retrieval.

TABLE 1
OBSERVING LOG AND $H_2O$ PRODUCTION RATES

| UT date | NIRSPEC Setting | $R_h$ [AU] | $dR_h/dt$ [km s$^{-1}$] | $\Delta$ [AU] | $d\Delta/dt$ [km s$^{-1}$] | $T_{int}$ [min] | $Q(H_2O)$ ($10^{26}$ s$^{-1}$) |
|---|---|---|---|---|---|---|---|
| 28 Nov 2004 | KL1 | 1.493 | −15.2 | 0.654 | −21.7 | 8 | 1535 ± 69 |
|  | KL2 | 1.493 | −15.2 | 0.655 | −21.8 | 20 | 1435 ± 124 |
| 29 Nov 2004 | MWA | 1.484 | −15.0 | 0.642 | −21.5 | 6 | 1253 ± 153 |
| 19 Jan 2005 | KL1 | 1.208 | −2.0 | 0.393 | 10.9 | 8 | 2727 ± 114 |
|  | KL2 | 1.208 | −2.0 | 0.394 | 11.0 | 8 | 2755 ± 132 |

NOTE. – $R_h$, $dR_h/dt$, $\Delta$, and $d\Delta/dt$ are respectively heliocentric distance, heliocentric radial velocity, geocentric distance, and topocentric line-of-sight velocity of C/2004 Q2; $T_{int}$ is total integration time on source, and $Q(H_2O)$ is the water production rate, as described in §3

TABLE 2a
ORGANIC VOLATILE COMPOSITION OF COMET C/2004 Q2 (MACHHOLZ)

| NIRSPEC Setting | Parent Molecule | $T_{rot}$ [a] [K] | Q [b,c] ($10^{26}$ s$^{-1}$) | Mixing Ratio [%] |
|---|---|---|---|---|
| \multicolumn{5}{c}{2004 Nov 28, $R_h$ = 1.493 AU, $\Delta$ = 0.65 AU} | | | | |
| KL1 | $H_2O$ | (86) | 1535 ± 55 (69) | 100 |
| | $C_2H_6$ | (86) | 8.62 ± 0.35 | 0.56 ± 0.03 |
| | $CH_3OH$ | (86) | 31.11 ± 1.21 | 2.03 ± 0.11 |
| | $CH_4$ | (86) | 19.21 ± 1.65 | 1.25 ± 0.12 |
| KL2 | $H_2O$ | 86 ± 4 | 1435 ± 73 (124) | 100 |
| | HCN | 76 ± 9 | 2.02 ± 0.20 | 0.14 ± 0.02 |
| | $NH_3$ | (86) | < 6.81 [d] | < 0.47 [d] |
| | $C_2H_2$ | (86) | < 0.91 [d] | < 0.06 [d] |
| | $H_2CO$ | (86) | 1.45 ± 0.45 | 0.10 ± 0.03 |
| | $CH_4$ | (86) | 18.08 ± 0.36 | 1.26 ± 0.10 |
| \multicolumn{5}{c}{2004 Nov 29, $R_h$ = 1.484 AU, $\Delta$ = 0.64 AU} | | | | |
| MWA | $H_2O$ | (100) | 1253 ± 112 (153) | 100 |
| | CO | $97^{+18}_{-17}$ | 63.46 ± 2.77 | 5.07 ± 0.51 |
| \multicolumn{5}{c}{2005 Jan 19, $R_h$ = 1.208 AU, $\Delta$ = 0.39 AU} | | | | |
| KL1 | $H_2O$ | (93) | 2727 ± 70 (114) | 100 |
| | $C_2H_6$ | (93) | 14.89 ± 0.85 | 0.55 ± 0.04 |
| | $CH_3OH$ | (93) | 61.97 ± 2.12 | 2.27 ± 0.12 |
| | $CH_4$ | (93) | 39.05 ± 2.41 | 1.43 ± 0.10 |
| KL2 | $H_2O$ | 93 ± 2 | 2755 ± 75 (132) | 100 |
| | HCN | 76 ± 2 | 4.12 ± 0.07 | 0.15 ± 0.01 |
| | HCN | (93) | 4.51 ± 0.18 | 0.16 ± 0.01 |
| | $NH_3$ | (76) | 8.26 ± 1.20 | 0.30 ± 0.04 |
| | $NH_3$ | (93) | 10.22 ± 1.48 | 0.37 ± 0.06 |
| | $C_2H_2$ | (76) | 2.31 ± 0.12 | 0.08 ± 0.01 |
| | $C_2H_2$ | (93) | 2.52 ± 0.13 | 0.09 ± 0.01 |
| | $H_2CO$ | (93) | 3.38 ± 0.35 | 0.12 ± 0.02 |
| | $CH_4$ | (93) | 42.50 ± 1.20 | 1.54 ± 0.06 |

[a] Rotational temperature. Values in parenthesis are assumed.

[b] Errors in production rate include both photon noise and line-by-line deviation between modeled and observed intensities (see Bonev et al. 2007, Dello Russo et al. 2004, Bonev 2005). For Q($H_2O$), the additional error in parenthesis includes also the uncertainty associated with slit losses caused primarily by atmospheric seeing (see DiSanti et al. 2006, Bonev et al. 2006). Slit losses do not affect mixing ratios among parent species measured within the same NIRSPEC setting, but should be considered in comparison between $H_2O$ production rates measured from different settings and/or on different dates.

[c] See also Table 3 and §4 for methane results on 2005 Jan. 19.

[d] Upper limits are 3σ.

TABLE 2b

WEIGHTED MEAN MIXING RATIOS FOR SPECIES DETECTED ON BOTH DATES
(NOV. 28, 2004 AND JAN. 19, 2005)

| Volatile | HCN | $CH_4$ | $C_2H_6$ | $H_2CO$ | $CH_3OH$ |
|---|---|---|---|---|---|
| Mixing ratio $X/H_2O$ [%] | $0.15^{+0.01}_{-0.02}$ | $1.46 \pm 0.08$ | $0.56 \pm 0.02$ | $0.11 \pm 0.03$ | $2.14 \pm 0.12$ |

TABLE 3
RELATIVE ABUNDANCES OF $CH_4$ AND $CH_3D$ IN C/2004 Q2 (MACHHOLZ)
(JANUARY 19, 2005)

| NIRSPEC Setting | Isotope | $T_{rot}$ [K] | Q ($10^{26}$ s$^{-1}$) | Mixing Ratio [%] X/$CH_4$ | Mixing Ratio [%] X/$H_2O$ |
|---|---|---|---|---|---|
| KL1 | $CH_4$ | (93) | 39.05 ± 2.41 | 100 | 1.43 ± 0.10 |
| KL2 | $CH_4$ | (93) | 42.50 ± 1.20 | 100 | 1.54 ± 0.06 |
| KL2 | $CH_3D$ | (93) | 0.025 ± 0.284 [a] | < 2.0 [a] | |

NOTE. – $T_{rot}$ and Q are defined as in Table 2a.

[a] The error in Q($CH_3D$) ($\sigma_Q$) exceeds the formal measured value, i.e., $CH_3D$ was not detected. The upper limit for the $CH_3D$ mixing ratio corresponds to 3 x $\sigma_Q$($CH_3D$)/Q($CH_4$) (KL2 setting). See text (§ 4) for a discussion of the factors affecting the sensitivity of the resulting upper limit.

TABLE 4
RELATIVE FLUXES BETWEEN THE R0 AND R1 LINES OF CH$_4$ ($\nu_3$ BAND):
SENSITIVITY TO ASSUMED SPIN TEMPERATURE

| UT Date | Measured Flux Ratio (R0/R1) | Model $T_{rot}$ [K] | Predicted Flux Ratio (R0/R1) | | |
|---|---|---|---|---|---|
| | | | Statistical equilibrium | $T_{spin} = 38$ K | $T_{spin} = 30$ K |
| 28 Nov. 2004 | 1.126 ± 0.043 | 65 | 1.239 | 1.290 | 1.401 |
| | | 86 | 1.172 | 1.219 | 1.323 |
| | | 90 | 1.163 | 1.209 | 1.313 |
| 19 Jan. 2005 | 1.068 ± 0.045 | 70 | 1.220 | 1.269 | 1.378 |
| | | 93 | 1.157 | 1.203 | 1.307 |
| | | 110 | 1.128 | 1.173 | 1.275 |

Notes. – All flux ratios are corrected for telluric transmittance. The CH$_4$ rotational temperature ($T_{rot}$) cannot be constrained from our spectra and therefore is used as a free parameter. Rotational temperatures of 86 K (November) and 93 K (January) were found for H$_2$O (Table 2a).